\documentclass[11pt,twocolumn]{article}
\usepackage[utf8]{inputenc}
\usepackage{hyperref}
\usepackage{amsmath}
\usepackage{amsfonts}
\usepackage{amssymb}
\usepackage{commath}
\usepackage{bm}
\usepackage{xcolor}
\usepackage{graphicx}
\usepackage{float} 
\usepackage{authblk} 
\usepackage[backend=bibtex,maxnames=3,url=false,
  style=numeric,isbn=false,giveninits=true,eprint=false,
  sorting=none,doi=false,date=year]{biblatex}
\usepackage[a4paper, total={7.2in, 9.7in}]{geometry}

\DeclareUnicodeCharacter{03B3}{$\gamma$}

\DeclareMathOperator{\argmax}{argmax}

\newtheorem{theorem}{Theorem}[section]
\newtheorem{remark1}[theorem]{Remark}
\newenvironment{remark}{\begin{remark1} \rm}{\end{remark1}}

\title{Methods for Cryo-EM Single Particle Reconstruction of Macromolecules having Continuous Heterogeneity}
\author{Bogdan Toader, Fred J. Sigworth and Roy R. Lederman}
\date{Yale University}

\bibliography{references}

\begin{document}

\maketitle

\abstract{
    Macromolecules change their shape (conformation) in the process of carrying out their functions. The imaging by cryo-electron microscopy of rapidly-frozen, individual copies of macromolecules (single particles) is a powerful and general approach to understanding the motions and energy landscapes of macromolecules. Widely-used computational methods already allow the recovery of a few distinct conformations from heterogeneous single-particle samples, but the treatment of complex forms of heterogeneity such as the continuum of possible transitory states and flexible regions remains largely an open problem.
    In recent years there has been a surge of new approaches for treating the more general problem of continuous heterogeneity.
    This paper surveys the current state of the art in this area.   
%
}

\section{Introduction}

Over the past few years, the combination of cryo-electron microscopy (cryo-EM) imaging and single-particle analysis has been established 
as the method of choice for studying the structure of large protein 
complexes at atomic or near-atomic resolution ~\cite{noauthor_method_2016}. 
Its recent success has been enabled by advances in detector technology, sample 
preparation techniques and the availability of advanced image processing 
software packages. The two other major structure-determinations methods are X-ray crystallography, which requires the sample to 
be crystallized, and NMR, which is useful only with relatively small proteins. 

Cryo-EM single-particle analysis (we will denote this simply as cryo-EM) involves the imaging of individual copies (called particles) of a macromolecular structure. Through the computational processing of $10^4$ to $10^6$ of such particle images, 3D density maps can be obtained through \textit{single-particle reconstruction}. To density maps of sufficient resolution, atomic structures can then be fitted, with $\sim 4$ \AA being the worst resolution for successful ab-initio fitting. Because the macromolecules are suspended in solution before they are rapidly frozen, particle images are likely to reflect a more native conformation, but also may contain frozen instances of flexibility or conformational variation.
Therefore, one of the promises of cryo-EM is that researchers will 
be able to construct a complete picture of all the possible conformations
of the imaged structures.

Conformational changes are key to the function of many macromolecular machines. The molecular motors dynein and kinesin undergo cyclical changes as a chemical reaction (hydrolysis of ATP) drives a mechanical stepping motion that moves cargo along a microtubule filament. Glucose transporters allow cells to take up this nutrient through a conformational cycle that enforces the transport of one or two Na$^+$ ions with each glucose molecule. DNA replication is carried out by the replisome, a large combination of molecular machines that, in stepwise fashions, unwind the double-stranded DNA and synthesize new complementary DNA strands.

While existing methods excel at reconstructing clearly-defined discrete conformations from cryo-EM data, the problem of reconstructing continuously-varying conformations of a macromolecule
is where most state of the art methods fall short. This is the \textit{continuous heterogeneity reconstruction} problem.
Fortunately, there has been a surge in the effort devoted to developing 
computational tools for continuous heterogeneity reconstruction, and this is 
the focus of this survey article. Our aim is to highlight the defining 
characteristics of each method and the conceptual overlaps and differences between them.

We note that there are many ideas and approaches for continuous heterogeneity and many papers introduce multiple new ideas and combine multiple approaches. 
For brevity and clarity, we cluster together different works and omit important implementation details. 
This paper covers the conceptual families of ideas and does not make specific recommendations about software to use.  
Some of the work surveyed is theoretical or less accessible to the user. Where available, we included links to some of the software that may be more accessible.

\subsection{Complementary Surveys}

There are a number of recent surveys covering other aspects of the cryo-EM pipeline. 

A general review of the computational challenges and the main components in the analysis pipeline are available in \cite{bendory_single-particle_2020}. A comprehensive description of the mathematical aspects of the problem, focusing on homogeneous reconstruction and validation, are available in \cite{singer_computational_2020}.

Discrete heterogeneity is discussed in \cite{jonic_cryo-electron_2016}.
A survey of earlier work on continuous heterogeneity, with a comprehensive survey of normal mode analysis (Section \ref{sec:nma}) is available in \cite{sorzano_survey_2019}.
In~\cite{cossio_likelihood-based_2018}, the focus is on the interpretation
of the energy landscape resulting from the heterogeneity analysis
using likelihood-based methods.
An up-to-date overview of the full cryo-EM pipeline, from sample preparation to 
the latest reconstruction methods, including time-resolved cryo-EM is available in \cite{devore_probing_2022}.

The recent reviews~\cite{wu_machine_2022} and~\cite{donnat_deep_2022}
focus on machine learning approaches to cryo-EM. Specifically,
the former gives an overview of machine learning algorithms used
in each step in the cryo-EM pipe-line, from pre-processing and
particle picking to 3D reconstruction and post-processing, while
the latter is a thorough survey of deep generative modeling 
techniques for 3D reconstruction.

In this review, we summarize the state of the art methods for analyzing continuous heterogeneity in cryo-EM. Our aim is
to sort the main families of ideas in the area and convey some of the main ideas of each technique.

\subsection{Outline of the Paper}

We first discuss a simplified image formation model that most cryo-EM 
reconstruction methods assume, as well as discrete heterogeneous 
reconstruction and multi-body refinement, in Section~\ref{sec: prelims}.

In Section~\ref{sec:manifoldEM} and Section~\ref{sec:manifoldvol},
we describe manifold learning approaches to continuous heterogeneous
reconstruction, specifically applied to particle images and reconstructed volumes
respectively. In Section~\ref{sec:linearvol} and Section~\ref{sec:nma},
we discuss linear models based on
covariance estimation and normal mode analysis respectively.

In Section~\ref{sec:hyper}, we discuss nonlinear models for continuous
heterogeneous reconstruction which conceptually fit into the ``hyper-molecules'' framework, including traditional and deep learning algorithms.
In Section~\ref{sec:matching dist}, we discuss inference methods that bypass 
the traditional treatment of latent variables.

Finally, we conclude the article with a discussion in Section~\ref{sec:discussion}.

\section{Preliminaries}
\label{sec: prelims}

\subsection{Image Formation Model and Homogeneous Reconstruction}
\label{sec: image formation}

In this section, we discuss a simplified model for image formation that
is the basis of all the approaches in this survey. 
For simplicity, we begin this discussion with a model for the tomographic projections of a single conformation in the \textit{homogeneous} case.
While each specific 
method may be variations of this model and contain additional details,
this model provides a useful baseline.

Let $V(\mathbf{r})$ represent the electrostatic 
potential of the molecule of interest in a specific conformation
at location $\mathbf{r} \in \mathbb{R}^{3}$. 
Throughout this survey, we will use the terms \textit{volume} and \textit{density} 
interchangeably to refer to $V$.

The \textit{particle images} $I_i$ ($i = 1, \ldots, M$)
are given by the following linear model of the image formation process:
\begin{equation}
    I_i = (C_i \circ T_i \circ P \circ R_i)V + \eta_i,
    \label{eq: image formation model}
\end{equation}
for $i = 1, \ldots, M$, where $R_i$ is a 3D rotation operator 
corresponding to the orientation of the volume $V$, $P$ is 
the 2D projection operator and $T_i$ is a 2D translation operator
corresponding to the offset of the projected volume with respect
to the center of the image.
Both $R_i$ and $T_i$ are specific to each particle image $I_i$
and are unknown in advance. 
$C_i$ is the  \textit{contrast transfer function} (CTF) operator applied to 
the projected image; the operator can be different for each particle image, and in our simplified discussion let us assume that $C_i$ is known.

For simplicity, we assume that the particle images are cropped in advance from the larger micrograph produced by the microscope. 

Lastly, Gaussian noise $\eta_i$ is applied to each image. In reality, the
noise introduced in cryo-EM images is shot noise~\cite{glaeser_single-particle_2021},
but the Gaussian assumption is commonly used in the reconstruction literature.
While the details of the noise model depend on the specific reconstruction 
method, it is sufficient for our discussion to assume that the noise parameters are estimated in advance.

\begin{remark}
    Cryo-EM algorithms make extensive use of the Fourier transform of images and density functions for efficient representation and computation. While the distinction between representations and operations in the Fourier domain and in the spatial domain is very important in implementations, it is not specific to the study of heterogeneity. For brevity, we omit the detailed discussion of Fourier vs. spatial domain implementations in different algorithms and focus on the main ideas that are specific to the study of heterogeneity.
\end{remark}

In the case of homogeneous reconstruction, the usual approach is to solve
the maximum-a-posteriori (MAP) problem given the set of particle images $\{I_i\}$:
\begin{equation}
    \argmax_{V} \ln p(V | \{I_i\}),
\end{equation}
where the log-posterior distribution is given by:
\begin{align}
    \ln~&p(V | \{I_i\}) = 
    \nonumber \\
    &\sum_{i=1}^M \ln 
    \int_{\phi_i} p(I_i | V, \phi_i) p(\phi_i) \dif \phi_i
    + \ln p(V).
    \label{eq: map}
\end{align}
Here, the likelihood function $p(I_i | V, \phi_i)$ is Gaussian
and determined by the image formation model~\eqref{eq: image formation model},
and the rotation and the translation $R_i, T_i$ of each particle image are 
paired into one pose variable $\phi_i$ whose assumed joint distribution 
is denoted by $p(\phi_i)$. From a Bayesian point of view, $p(\phi_i)$ 
and $p(V)$ play the role of priors (of the pose and the volume respectively), 
and in some methods (e.g., RELION), they are adjusted iteratively 
during reconstruction.

\subsection{Discrete Heterogeneity}
\label{sec:discrete}

The traditional approach to the heterogeneity problem is to perform 3D classification,
where each particle is assigned to one of a small number of different, optimized reference
volumes (in some algorithms, a probability of assignment to each references volume is considered). 
This works best when the underlying macromolecule has a small number of discrete states, and these are 
distinguishable in the set of single-particle images.

The main assumption here is that the reference volumes obtained are a good 
representation of the different conformations of the macromolecule, which requires
the removal of the outliers before reconstruction and that each class contains
enough particle images for the reconstruction to be possible. Usually
the number of reference volumes, or classes, is specified in advance, but
it can change during optimization if some classes are not very
populated~\cite{kimanius_new_2021}.

First introduced in~\cite{scheres_disentangling_2007}, this method consists of
maximum likelihood or maximum-a-posteriori optimization for $K$ different 
volumes, or $K$ \textit{classes}, 
$\{V_1(\mathbf{r}), \ldots, V_K(\mathbf{r})\}$.
Then, the log posterior distribution~\eqref{eq: map} becomes:
\begin{align}
    \ln~&p(\{V_k\} | \{I_i\}) = 
    \nonumber \\
    &\sum_{i=1}^M \ln \left( \sum_{k=1}^K  
        \int_{\phi_i} p(I_i | k, \{V_k\}, \phi_i) p(k, \phi_i | \{V_k\}) \dif \phi_i
    \right)
    \nonumber \\
    +&\sum_{k=1}^K \ln p(V_k).
    \label{eq: map K}
\end{align}

The $K$ volumes are usually initialized as 
low dimensional reconstructions from $K$ randomly
drawn subsets of the particle images,
and the optimization can be performed, for example,
with the expectation-maximization algorithm
~\cite{scheres_disentangling_2007, scheres_relion_2012,lyumkis_likelihood-based_2013}
or with stochastic gradient descent or its variants 
\cite{punjani_cryosparc_2017, kimanius_new_2021}.\footnote{
    For software, see for example the RELION \url{https://relion.readthedocs.io} 
    or CryoSPARC \url{https://cryosparc.com} packages. 
    \label{fn:relioncryosparc}
}

\subsection{Multi-body Extension of Traditional Analysis} 
\label{sec:multibody}

A step up in complexity is the multi-body 
refinement~\cite{bai_sampling_2015,ilca_localized_2015,nakane_characterisation_2018}\footnote{
    For software that performs multi-body analysis, see RELION 
    in footnote~\ref{fn:relioncryosparc}.
}.
Here, the volume is assumed to consist of a small number of rigid components 
that move relative to each other and are identical across the particle images. 

This method starts with applying standard homogeneous reconstruction techniques
to obtain a consensus reconstruction of the bulk of the volume.  
Then, the user specifies a set of 3D masks that define regions of the volume 
containing individual components that move with respect to the bulk. 
Each of these components is treated like a rigid body. 
Lastly, independent homogeneous reconstruction is performed for each 
component using the particle images where, for each component reconstruction, 
the CTF-affected 2D projections of the other components have been subtracted 
from the particle images. The subtraction is done either before the separate 
components reconstruction step~\cite{bai_sampling_2015,ilca_localized_2015} 
or during the reconstruction process in an iterative 
fashion~\cite{nakane_characterisation_2018}. 
In the latter approach, the relative orientation of each component is refined
for each particle image and updated at every iteration before subtraction, 
which leads to improved results.

One obvious disadvantage of this approach is that it relies on the user's
previous knowledge of the structural domains of the molecule and, 
therefore, it is susceptible to human bias. In addition, multi-body refinement
is limited to rigid variability and has difficulty at the interface between the
moving components.

\subsection{Conformation Space}

\begin{figure*}[h]
    \centering
    \includegraphics[width=0.6\textwidth]{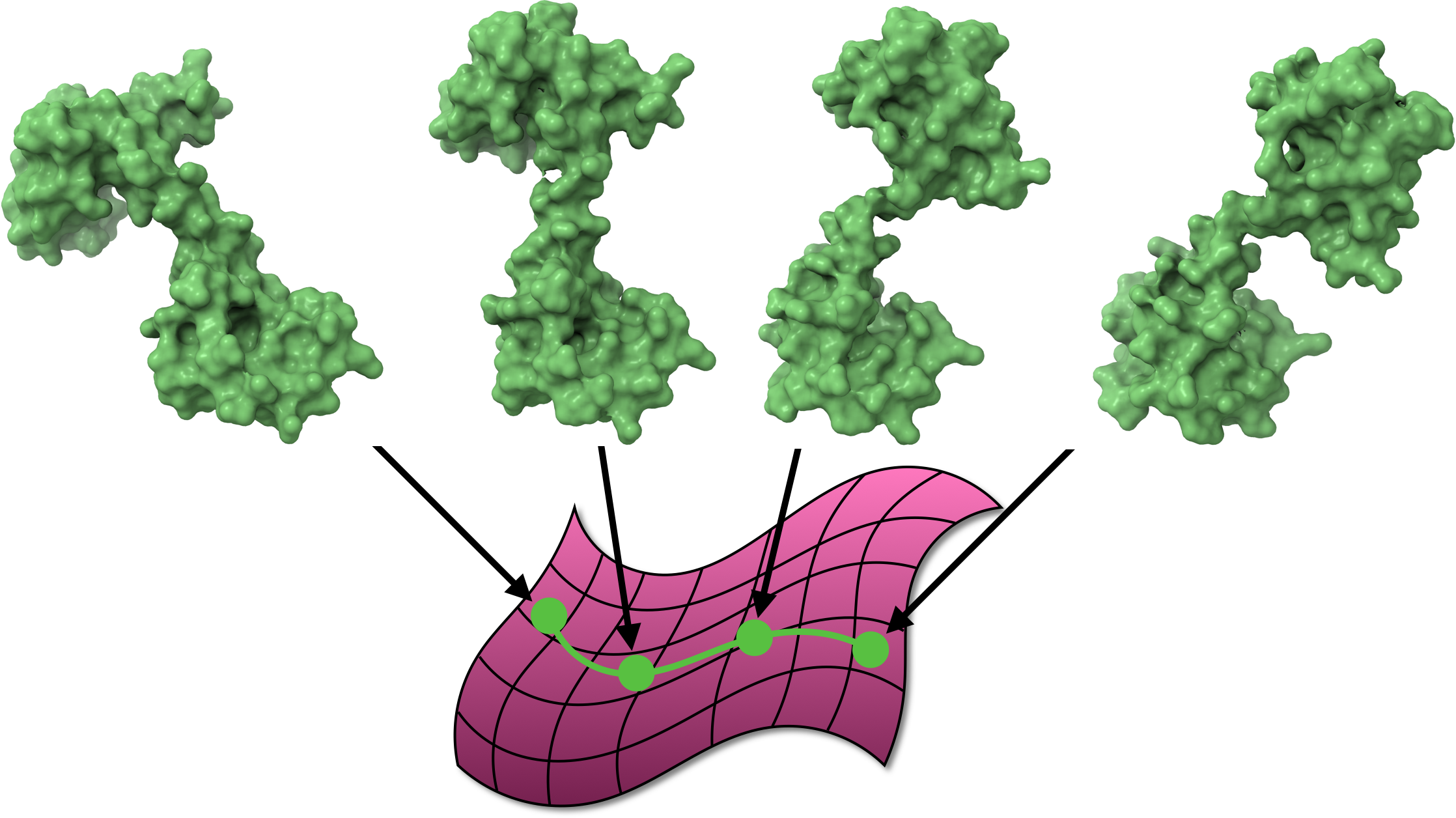}
    \caption{Examples of conformations of the calmodulin protein
        along a trajectory on its conformation manifold (illustration).
        Volumes obtained with UCSF ChimeraX version 1.4 (2022-06-03)
        using its \textit{morph} function and the structures from
        the Protein Data Bank with entryIDs 1CFD and 3CLN.}
    \label{fig:conformations}
\end{figure*}

Given a set of particle images, the objective of traditional homogeneous 
reconstruction is to recover a single volume which, according to the forward
model~\eqref{eq: image formation model} (or a variation of it), generated
the particle images. 
When there is conformational heterogeneity that a single homogeneous reconstruction model cannot capture, 
homogeneous reconstruction algorithms recover some average structure (or even worse corrupted structures)
Similarly, the main underlying assumption in discrete heterogeneous 
reconstruction is that the particles cluster around a small number of 
distinct dominating structures, which the reconstruction process
aims to recover.

In both cases, an important step of the data pre-processing is the removal of 
outliers that do not fit with any of the structures that one wants to
recover. Moreover, information from particles on the boundary
between different ``discrete'' structures, which may represent rare or less
likely conformations, is lost when they are assigned 
to one of the volumes or classified as an outlier.

The strength of cryo-EM for structure determination of biological samples
is that it is able to capture the sample and its conformations without being constrained to a crystal lattice. 
Cryo-EM observes individual molecules rather than crystal ensemble averages. 
Therefore, avoiding averaging of these conformations
through reconstruction of one or several discrete volumes is a key 
motivation for continuous heterogeneous reconstruction.

While the output of homogeneous or discrete heterogeneous reconstruction
consists of a small number of discrete volumes, the output of continuous
heterogeneity analysis should capture the full space of conformations
of the molecule. This often takes the form of a \textit{low-dimensional manifold}. 
The specifics vary significantly between different techniques. 
Throughout this survey, we will refer to this manifold
as the \textit{conformation space} or \textit{latent space} representation.
Given such a manifold
and a coordinate vector on it, one should be able to recover a volume 
corresponding to the conformation at those coordinates. 
In Figure~\ref{fig:conformations}, we illustrate examples of volumes along
a trajectory in the conformation space for the calmodulin protein.

\section{Manifold Learning on Particle Images }\label{sec:manifoldEM}

One of the first ideas for characterizing continuous heterogeneity in cryo-EM was based on the observation that a continuous space of volumes could be represented mathematically as a manifold of volumes; similarly, the space of particle images, which are the projections of different molecules 
in different viewing directions, 
form a manifold of images \cite{schwander_mapping_2010}. 
Manifold learning techniques are usually based on organizing observations (particle images) based on a norm of the difference between every two observations. The study of cryo-EM data as a manifold of particle images presents several challenges. For example, in the simplest form of manifold learning, each variable such as the viewing direction, in-plane rotations, in-plane translations, and CTF, introduces additional dimensions to the problem, making the analysis impractical even before considering the heterogeneity. Furthermore, the technical properties of the difference between particle images lead to mathematical difficulty in identifying the observed manifold with the product $SO(3) \times \tau$ of viewing directions and heterogeneity space. Finally, the high level of noise in cryo-EM make the difference between particle images noisy.
A sequence of followup work introduced additional ideas to make the analysis practical~\cite{dashti_trajectories_2014, schwander_conformations_2014, frank_continuous_2016, maji_propagation_2020, dashti_retrieving_2020, seitz_recovery_2022}\footnote{
    For the ManifoldEM/ESPER software, see the code repositories:
    \url{https://github.com/GMashayekhi/ManifoldEM_Matlab}, \\ 
    \url{https://github.com/evanseitz/ManifoldEM_Python}, \\
    \url{https://github.com/evanseitz/cryoEM_ESPER}.

}
by assuming known viewing directions and analyzing the manifold of conformations as viewed from each direction using the diffusion map algorithm \cite{coifman_diffusion_2006}; the maps are then aligned across all different directions either by using 
nonlinear Laplacian spectral analysis (NLSA)~\cite{dashti_retrieving_2020} or a novel 
algorithm ESPER~\cite{seitz_recovery_2022}.

Experiments performed using simulated data in~\cite{seitz_recovery_2022} show that the conformation manifold is not constructed well for certain projection directions where the range of the conformation variability is small. The manifold embeddings are also affected negatively by insufficient samples for each state in the range and low SNR, effects that are equally visible for both PCA and Diffusion Maps. To circumvent these issues, ESPER \cite{seitz_recovery_2022} rotates the manifolds' coordinates (\textit{eigenfunction realignment}) so that each dimension of the heterogeneity variable is captured in a separate eigenbasis of the embedding. 
Then, a subspace partitioning procedure is performed to assign particles to each state along the trajectory representing the continuous heterogeneity and produce a 2D movie for each viewing direction. 
The direction of each movie is determined using optical flow and belief propagation
using the method in~\cite{maji_propagation_2020}, after which the individual states 
along the trajectory can be reconstructed using standard homogeneous 
reconstruction methods.

The manifold approach in this section is characterized by sidestepping the modeling of the volumes themselves during the analysis. Instead, this approach aims to build a map of conformations 
for each viewing direction. 
Once the map is available, one selects particle images from many viewing directions of roughly the same conformation and reconstructs that conformation from the particle images using traditional reconstruction algorithms. 
In the following sections, we will discuss approaches that rely on 
explicit modeling of the volumes.

\section{Manifolds of Volumes}
\label{sec:manifoldvol}

Many of the challenges presented in the previous section can be avoided by applying manifold learning techniques directly to volumes, which is the approach we discuss in this section.
Clearly, cryo-EM experiments produce tomographic projections, not volumes. 
Therefore, the approach in this section of not self-contained in most cases.
One work on manifolds of volumes in~\cite{sanchez_sorzano_structmap_2016} assumes that the volumes are given.
A later method in~\cite{wu_visualizing_2022} generates the volumes by performing
3D classification of reconstructed volumes from a number of bootstrapped
subsets of the data.

Once the volumes are available,
 both methods propose a way of obtaining the low dimensional embedding
corresponding to the manifold of conformations. 
In StructMap\footnote{
    StructMap is distributed as part of the ContinuousFlex
    plugin for Scipion (see footnote~\ref{fn:hemnma}).
}~\cite{sanchez_sorzano_structmap_2016}, the volumes are aligned 
using rigid body alignment as well as flexible alignment using their 
normal mode representations (which will be discussed in Section~\ref{sec:nma}), 
resulting in a dissimilarity matrix containing pair-wise distances 
between volumes. Then, multi-dimensional scaling is applied to a distance 
matrix obtained from the dissimilarity matrix. 
In AlphaCryo4D\footnote{
    The AlphaCryo4D code can be found at \\
    \url{https://github.com/alphacryo4d/alphacryo4d/} and 
}~\cite{wu_visualizing_2022}, the low dimensional manifold embedding is 
obtained by first extracting features
from the volumes using an autoencoder neural network, and then applying
t-SNE~\cite{van_der_maaten_visualizing_2008} to the volumes and their features.

Once the low dimensional manifold of conformations is obtained, one can analyze
the conformational landscape, for example by performing clustering on the 
embedding, or high-resolution homogeneous refinement based on specific regions
of the manifold.

Since this approach relies on volumes that are produced by some other methods, in inherits some difficulties in generating such volumes from other methods. 
That being said, many of the methods in the remaining sections involve some components of dimensionality reduction or manifold learning on the latent conformation space, which are conceptually related to the methods in this section and the previous section.
In particular, a different take on manifold learning of volumes, presented in~\cite{moscovich_cryo-em_2020} combines manifold learning techniques with covariance estimations that we discuss in the next section.

\section{Principal Volumes and Linear Models for Volumes} \label{sec:linearvol}

The approaches in this section and the remaining sections incorporate generative models that describe the space of conformations directly. 
This is in contrast to Section~\ref{sec:manifoldEM}, which maps the conformation space without creating an internal representation of the space of possible volumes, and in contrast to Section~\ref{sec:manifoldvol}, where the manifold is not inherently a generative model (with some exceptions such as \cite{moscovich_cryo-em_2020} that integrates ideas from this section and the previous section).

One approach to representing the densities is based on linear combinations of volumes. 
In this representation, we have a reference volume, which we denote $X_0$, and $K$ principal volumes, which we denote by $\{X_k\}_{k=1}^K$.  
The density $V_m := V(m, \cdot)$ of a particular conformation is provided as a linear combination of the principal volumes, 
\begin{equation}\label{eq:linearvol}
    V_m \approx X_0 + \sum_{k=1}^K b^m_k X_k, 
\end{equation}
where the coefficient $b^m_k$ is the weight that the $k$-th principal volume receives in the $m$-th conformation; there is one conformation (and one set of coefficients) corresponding to each particle image. 
The coefficients $b^m_k$ are different for each conformation and can be positive or negative, 
and the principal volumes are often chosen to be orthogonal to each other, 
such that $\int X_k(x) X_n(x) dx = 0$ if $k \ne n$ (excluding $X_0$).
Once the principal volumes are obtained, the standard practice is to visualize each principal mode separately. The expression for the movie of the $k$-th principal mode as a function of ``time'' $\tau$ (negative or positive) is
\begin{equation}\label{eq:linearvol:modes}
    V^k(\tau) = X_0 + \tau X_k.
\end{equation}

If the volumes in different conformations were given (and aligned), 
a natural optimal choice of principal volumes $X_k$ is obtained by a standard 
Principal Component Analysis (PCA) of the densities. This idea was first 
introduced in~\cite{liu_estimation_1995} and further developed 
in~\cite{penczek_identifying_2011}, where the volumes are obtained by 
homogeneous reconstruction using resampled subsets of the particle images.

Remarkably, the authors of~\cite{tagare_directly_2015} and~\cite{katsevich_covariance_2015} demonstrate that it is possible
to estimate the 3D covariance matrix and principal volumes directly from the 2D particle images under mild assumptions.
This surprising fact is evident from the expressions for the covariance in Fourier domain, extending the Fourier slice theorem.
Specifically, in~\cite{katsevich_covariance_2015}, appropriate estimators 
$\tilde{X}^M$ and $\Sigma^M$ for $X_0$ and the covariance matrix $\Sigma_0$ 
are defined using least-squares optimization problems. These estimators
are shown to be consistent, i.e. they converge to $\mu_0$ and $\Sigma_0$ as 
the number of particle images $M$ tends to infinity. Finding $\mu_M$ and 
$\Sigma_M$ then involves solving two linear systems. 
The work in~\cite{liao_efficient_2015,anden_covariance_2015,anden_structural_2018}
improves the solvers, both in terms of scalability and generality.
The authors of~\cite{tagare_directly_2015} propose a probabilistic PCA approach,
where the principal volumes are estimated directly from the data without first 
computing the covariance matrix $\Sigma_0$. This method is improved 
in~\cite{melero2020continuous} and in~\cite{punjani_3d_2021} (implemented in 
CryoSPARC), which allow reconstruction of volumes of higher 
resolution\footnote{
    Code for the principal volumes method is available as part of the 
    ASPIRE software package \url{http://spr.math.princeton.edu},
    as well as in CryoSPARC with its 3D variability analysis functionality
    (link given in footnote~\ref{fn:relioncryosparc}).
}.

The interpretation of heterogeneity based on the modes in (\ref{eq:linearvol:modes}) has been demonstrated in several 
examples~\cite{liao_efficient_2015,melero2020continuous,punjani_3d_2021}.
However, as discussed in more detail in~\cite{sorzano_principal_2021}, the covariance method has limited applicability to high-resolution approximation of large continuous conformational variability.
The problem, in a nutshell, is that high-resolution large continuous conformational variability does not behave like Equation (\ref{eq:linearvol:modes}).
Instead, one would require a larger number of high-resolution principal volumes 
and to identify specific combinations of coefficients $b^m_k$ in Equation (\ref{eq:linearvol}) that represent valid conformations. 

A refined and potentially more interpretable analysis is proposed in \cite{moscovich_cryo-em_2020}.
Each particle image can be best explained as a tomographic projection of a particular volume that is a specific linear combination (equation (\ref{eq:linearvol})) with particular coefficients $b^m_k$. 
This translates each particle image to an approximate volume, which can then be used for manifold learning on volumes (Section \ref{sec:manifoldvol}). 
The algorithm in \cite{moscovich_cryo-em_2020} further extends the manifold-learning approach to produce refined basis volumes, called \textit{spectral volumes}, that are compatible with the recovered manifold.

A key limitation of this approach to the analysis of heterogeneity is the prerequisite of known viewing directions for each particle image. Large conformational heterogeneity makes it challenging to define and compute consistent viewing directions.

\section{Normal Modes and Predefined Spaces of Conformations}\label{sec:nma}

In this section, we discuss a direct representation of the positions of atoms (or pseudoatoms) in each conformation.
Let $\mathbf{r}_{l}=(x_{l}, y_{l}, z_{l})^\intercal$ be the position of the $l$-th atom (or a pseudoatom) in a molecule, and let $\mathbf{r} = (\mathbf{r}_{1}^\intercal, \mathbf{r}_{2}^\intercal, ...)^\intercal \in \mathbb{R}^{3N}$ be the concatenation of the coordinates of $N$ atoms. 

The structural variability of molecules can be simulated and studied computationally by Normal Mode Analysis (NMA)~\cite{cui_normal_2005}.
NMA produces a linearized model for small perturbations around a reference location of the atoms. 
    The best reference structure is an (already determined) atomic model of the molecule, where the way in which the
    protein is folded will distinguish
    rigid (e.g. hydrogen-bonded) from non-rigid regions. An alternative but much less powerful approach 
    treats a low-resolution envelope of the molecule (say, as determined by cryo-EM) as an elastic structure. In this 
    model-free NMA, larger pseudo-atoms are used for a coarse 
    representation of the molecule.
    
Analogously to Section \ref{sec:linearvol}, the positions of atoms $\mathbf{r}^n$ in any conformation 
can be approximated by a linear combination of the normal modes $\{\Delta_k\}_{k=1}^K$
added to the reference positions $\mathbf{r}^0$:
\begin{equation}\label{eq:linearatom}
    \mathbf{r}^m \approx \mathbf{r}^0 + \sum_{k=1}^K b^m_k \Delta_k
\end{equation}
where the coefficients $b^m_k$ determine the amplitude of the perturbation 
in the $m$-th conformation and each set of coefficients $\{b^m_k\}_{k=1}^K$ 
corresponds to a particle image. 
The coefficients $b^m_k$ are different for each conformation, and they can be positive or negative. 
While the normal modes $\{\Delta_k\}_{k=1}^K$ are determined by the 
physics of the molecule, the coefficients $\{b_k^m\}_{k=1}^K$
are chosen so that the conformations fit the cryo-EM particle images.

The starting point of NMA is defining a potential function of the energy
of the reference reconstruction, which is usually considered to be a low energy
conformation. 

There are two distinct approaches to obtain the reference structure:
a model-based approach, where the reference structure is 
obtained from the atomic model of the molecule of interest, and a 
model-free NMA, where larger pseudo-atoms are used for a coarser 
representation of the molecule (obtained, for example, from a volume
determined using cryo-EM).
The quality of the normal modes varies with the quality of the model.

Then, the normal modes are the eigenvectors of the Hessian of the
potential function, with the normal modes corresponding to the large eigenvalues
describing large, collective motions of atoms and the modes corresponding to small
eigenvalues describing more localized motions. 
Once the normal modes are approximated from an approximate model of the molecule,
cryo-EM comes into the heterogeneity analysis to find the that best combination of normal modes to
explain each particle image.

In~\cite{tirion_large_1996}, a simple potential function
that models the distances between pairs of close
atoms as harmonic oscillators is introduced and is shown to 
accurately describe collective motions of atoms in a molecule as well
as more complex and computationally expensive potential that explicitly
model bond lengths, bond angles and dihedral angles. A web server 
allowing a user to compute the normal modes of a protein based on this
idea is described in~\cite{suhre_elnemo_2004}, and the same idea enables
recent approaches like HEMNMA\footnote{
    HEMNMA and DeepHEMNMA are distributed as part of the 
    ContinuousFlex plugin for the Scipion software package 
    \url{https://github.com/scipion-em/scipion-em-continuousflex}. 
    \label{fn:hemnma}
}~\cite{jin_iterative_2014} to scale to full cryo-EM datasets.

In early work in the context of cryo-EM, NMA has been used 
for flexible fitting of high resolution structures from X-ray
crystallography to low-resolution maps from cryo-EM~\cite{tama_normal_2004}
and for finding new conformations of a previously determined 
structure~\cite{brink_experimental_2004}. These ideas are taken further
in HEMNMA~\cite{jin_iterative_2014}, a method
for computing the conformational space of a reference structure
by fitting the normal mode coefficients $b^m_k$ 
that best explain each observed particle image in a cryo-EM dataset.
This works by iteratively refining the normal mode coefficients 
and pose coordinates (rotation angles and translations) for each particle image.

In~\cite{vuillemot_combined_2021} and~\cite{vuillemot_nmmd_2022},
global collective motions described by NMA are combined with local
atomic displacements given by Hamiltonian Monte Carlo sampling 
or molecular dynamics simulations. In DeepHEMNMA~\cite{hamitouche_deephemnma_2022}, 
a deep neural network is trained on particle images,
normal mode coefficients and pose parameters obtained from HEMNMA for a subset
of the data, which then outputs the corresponding normal mode coefficients
and pose parameters for the remaining data.
In~\cite{nashed_heterogeneous_2022}, an unsupervised 
learning approach is used to estimate normal mode coefficients and in-plane 
rotations, while in~\cite{woollard_physics_2022}, the coefficient of 
one normal mode is sampled using a deep encoder neural network together
with the CTF defocus and in-plane orientation of the particle.

Subsets of particle images that have similar values of coefficients are 
assumed to have the same conformations; such subsets can be used to reconstruct 
the volume directly from the images, or the conformations can be better 
visualized in a lower dimensional space by performing PCA on the normal
mode coefficients.
An important prerequisite for the algorithms is the reference $X_0$ and its atomic model, 
which is used to compute the normal modes $\{\Delta_k\}_{k=1}^K$.

WarpCraft~\cite{schilbach_structures_2017} models continuous heterogeneity
by using normal mode analysis to combine different components of the volume, 
which avoids the downsides of multi-body refinement of rigid components.

A comprehensive discussion of NMA is available in~\cite{cui_normal_2005} and in the 
context of cryo-EM in~\cite{sorzano_survey_2019}.

There are other approaches based on the physics of the molecules to provide conceptually related, but very different conformation spaces. 
Specifically, in~\cite{bonomi_simultaneous_2018}, molecular dynamics 
simulations are biased with cryo-EM data by sampling a posterior determined 
by an energy function that combines a standard molecular dynamics energy term 
and one that takes into account the cryo-EM data, 
and in~\cite{giraldo-barreto_bayesian_2021}, CryoBIFE uses a Bayesian approach 
to extract the free energy profile of a molecule directly from cryo-EM images 
together with its uncertainty.

\begin{remark}
    We note that in most normal mode analysis work described in the current section the space of conformations is predetermined by a fixed 
    the normal modes expressed in Equation \eqref{eq:linearatom}. 
    These normal modes are not inherently based on cryo-EM data, but rather on some external predefined model of the molecule (although that model might be derived from some reference homogeneous model reconstructed from cryo-EM data).
\end{remark}

\section{Nonlinear Models: Hyper-Molecules}
\label{sec:hyper}

In Sections \ref{sec:linearvol} and \ref{sec:nma} we discussed linear models, and in Section \ref{sec:manifoldEM} we discussed a nonlinear approach based on manifolds of two-dimensional tomographic projections. In this section we discuss nonlinear models for the volume $V$. 
The general formulation of nonlinear models in cryo-EM, laid out in \cite{lederman_continuously_2017,lederman_hyper-molecules_2020}, models all the conformations of the molecule in one {\em hyper-molecule} 
function $V(\bm \tau, \mathbf{r})$, 
where $\bm \tau$ is the conformation variable (position in the conformation space)
and $\mathbf{r}$ is a point in space (or a frequency).
In the simplest case, $\bm \tau := \tau$ is a scalar and the function 
$V(\tau,\mathbf{r})$ is analogous to a movie: if we fix a point $\tau$ 
in ``time'' we obtain a single conformation $V(\tau, \cdot)$ analogous 
to a single still frame in a video. In other words, $V(\tau, \mathbf{r})$ 
is a complete description of a continuum of conformations. 
More generally, $\bm \tau$ can be a high dimensional vector capturing complex 
continuous heterogeneity. For example, when $\bm \tau$ is two dimensional, 
one can imagine a planar map where each point represents a conformation. 

The hyper-molecule function $V(\bm \tau, \mathbf{r})$ has been implemented in 
many different ways in different works.

It can be argued that a continuous model can be approximated with a sufficiently large number of discrete samples, or a sufficiently large number of classes in 3D classification in cryo-EM. 
This would be analogous to the sequence of individual frames that represents a movie. 
However, the discrete 3D classes are analogous to an unordered collection of frames, which is not as interpretable as a movie. 
More importantly, the relation between adjacent frames in a video reduces the amount of information required to represent the movie (indeed, compressed videos are more efficient than a collection of images representing the same frames); this is loosely translates to a fewer particle images one would need in order to recover $V(\tau, \cdot)$ compared to number of particle images required to recover a very large number of discretized classes.
We revisit the motivation for continuous functions more formally in Section \ref{sec:contVsisc}.

\subsection{Orthogonal Basis Functions}
\label{sec:basis funcs}

The hyper-molecule $V(\bm \tau, \cdot)$ can be represented in different ways. 
The first classic harmonic analysis approach proposed in~\cite{lederman_continuously_2017,lederman_hyper-molecules_2020} uses linear combinations of orthogonal basis functions $P_k$:
\begin{equation}\label{eq:hypermolecule:orthogonal}
    V(\bm \tau, \mathbf{r}) = \sum_k a_{k} P_k(\bm \tau, \mathbf{r}).
\end{equation}
One practical implementation of (\ref{eq:hypermolecule:orthogonal}) is a product of three dimensional basis functions (such as prolate spheroidal functions~\cite{slepian1964prolate,lederman2017numerical,lederman_hyper-molecules_2020}) and one dimensional basis functions.

The principle was demonstrated to work using a stochastic gradient decent like algorithm on synthetic data in~\cite{lederman_continuously_2017}, 
and using a Markov Chain Monte Carlo (MCMC) algorithm and synthetic and real data in~\cite{lederman_hyper-molecules_2020}.
In both cases, the viewing direction and conformational state are not assumed to be known a-priori. 
While \cite{lederman_continuously_2017,lederman_hyper-molecules_2020} argue the hyper-molecules are the natural generalization of traditional 3D volumes to the case of continuous heterogeneity, it argued that it would become increasingly difficult to rigorously generalize the traditional expectation-maximization and branch-and-bound algorithms to high-dimensional latent spaces that involve viewing direction, translation and complex conformation variables. Therefore, they propose formal MCMC algorithms and speculate about possible use of variational approximations.

\begin{remark}
To contrast between equations (\ref{eq:hypermolecule:orthogonal}) and (\ref{eq:linearvol}), 
we note that the coefficients $a_k$ in (\ref{eq:hypermolecule:orthogonal}) are a shared part of the model,
whereas the coefficients $b^m_k$ in (\ref{eq:linearvol}) are different for each individual particle image.
\end{remark}

\begin{remark}
    In fact, the linear density model in covariance approach in Equation (\ref{eq:linearvol}) is a special case of nonlinear models.
    For example, one can define ${\bm \tau}$ to simply be the vector of coefficients ${\bm \tau}:= (b_1, b_2,.... b_K)$, so that
    $V_m({\bm r}) = V({\bm \tau^m}, {\bm r}) = X_0 + \sum_{k=1}^K b^m_k X_k$.
\end{remark}

\subsection{CryoDRGN}
\label{sec:cryodrgn}

\begin{figure*}[h]
    \centering
    \includegraphics[width=0.9\textwidth]{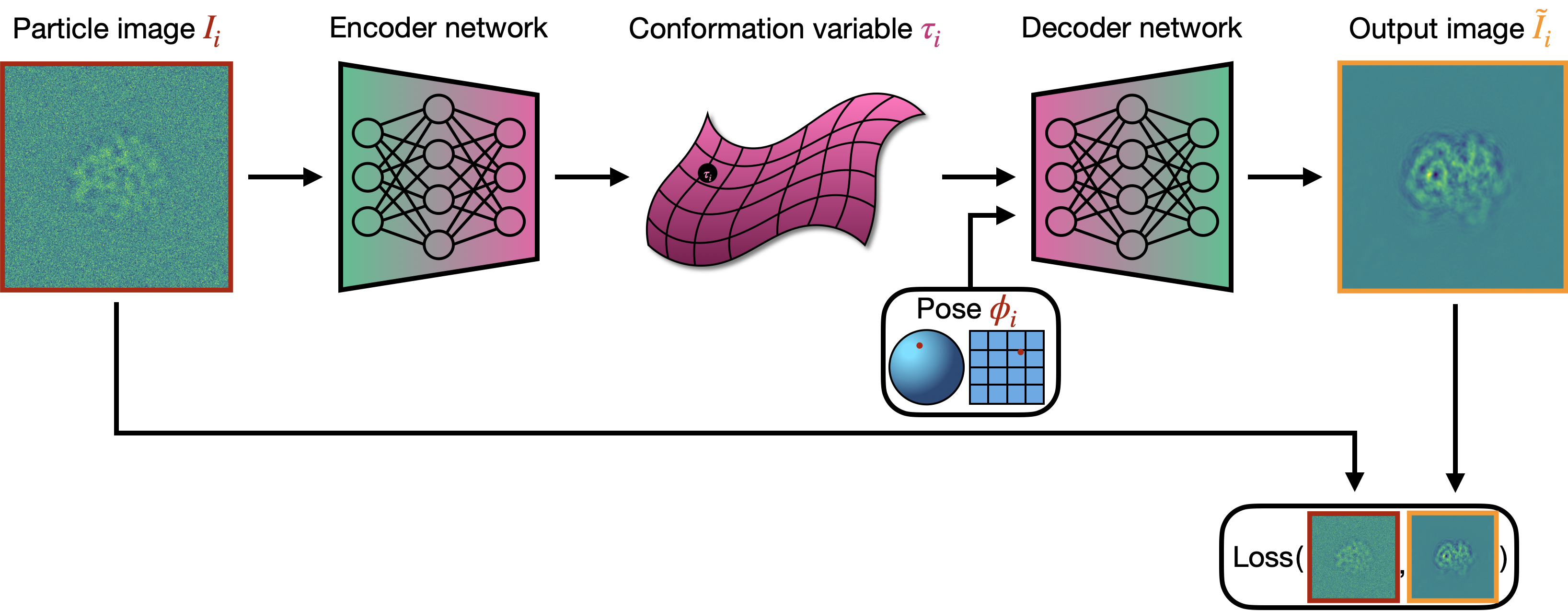}
    \caption{CryoDRGN architecture. In order to train the encoder and the 
        decoder networks, a particle image $I_i$ is given as input to the 
        encoder, which outputs a latent space representation (or conformation 
        variable) $\bm \tau_i$ of the particle image. Then, the decoder network takes 
        $\bm \tau_i$ and the pose variable $\phi_i$ as input and outputs a 
        projected image $\tilde{I}_i$ of the estimated volume. 
        The loss $\mathcal{L}(I_i, \tilde{I}_i)$ between the input 
        particle image and the output image is computed and the weights 
        of the networks are adjusted accordingly.
        The GMM and GMM with folding constraints models described in  
        Section~\ref{sec:gmm} and Section~\ref{sec:gmm folding} follow a 
        similar workflow, with the most significant difference being that 
        the decoder outputs the atom positions in the conformation 
        corresponding to $\bm \tau_i$, which are then used to generate 
        the output particle image $\tilde{I}_i$.
    }
    \label{fig:cryodrgn}
\end{figure*}

CryoDRGN\footnote{
    The CryoDRGN software is available at 
    \url{https://cb.csail.mit.edu/cb/cryodrgn}.
}~\cite{zhong_reconstructing_2019, zhong_cryodrgn_2021} introduce new 
ideas inspired by the success of Variational AutoEncoders 
(VAEs)~\cite{kingma_auto-encoding_2014} in other applications. 
VAEs have two components, an {\em encoder} which is optimized to provide approximations 
of the distribution of latent variables, and a {\em decoder}, which is optimized to 
reproduce particle images given the latent variables as inputs. 
Figure~\ref{fig:cryodrgn} is a schematic illustration of the cryoDRGN architecture. 
Many of the neural network based approaches discussed in the remainder of this section have an analogous architecture.

\begin{remark}
The use of a deep neural network in CryoDRGN and other methods below might create a perception 
that CryoDRGN is somehow pretrained on some dataset with many known structures; 
in fact, rather than being pre-trained, CryoDRGN fits the weights of the neural networks separately for each dataset to which it is applied.
\end{remark}

CryoDRGN's decoder is a hyper-molecule function $V(\bm \tau, \mathbf{r})$, implemented as a standard multilayer perceptron (MLP), with an interesting twist. 
The conformation latent variable $\bm \tau$ (typically 8 dimensional) feeds into the MLP in the traditional way.
The coordinates $\mathbf{r}=(r_1, r_2, r_3)$ (in the Fourier-Hartley domain) first go through positional encoding where each of the three coordinates is augmented; for example, the coordinate $r_1$ is augmented by the vector $\mathbf{\tilde{r}}_1$ with:
\begin{align}
    \mathbf{\tilde{r}}_1[2i] = \sin(r_1 N \pi (2/N)^{2i/N}), \\
    \mathbf{\tilde{r}}_1[2i+1] = \cos(r_1 N \pi (2/N)^{2i/N}), \\
\end{align}
where $i = 1,\ldots,N/2$, and $N$ is the size of the box containing the volume of interest.
The positional encoding modification to the MLP by augmenting the input 
vectors has been used in transformers and vision applications; it plays an 
important role in fitting high-frequency components of 
functions~\cite{tancik_fourier_2020}.

In order to produce a tomographic projection from a particular viewing direction and 
conformation $\bm \tau$, CryoDRGN computes a grid of points $r_{1,1},r_{1,2},...$ 
in the Fourier-Hartley domain and evaluates $V(\bm \tau, r_{1,1}), V(\bm \tau, r_{1,2}),...$ 
at these points. The grid of values is the Hartley transform of the particle image.

Previous algorithms store or compute for each particle image $I_i$ some form of 
explicit latent variables such as the viewing direction $R_i$ and 
conformation $\bm \tau_i$. For example, MCMC software stores these 
latent variables and software like RELION compares each particle image to projections of 
different volumes in different viewing directions.
Instead of storing the explicit $\bm \tau_i$ for each particle 
image $I_i$, CryoDRGN's encoder is optimized to take the image $I_i$ 
itself as input and compute a sample from $p(\bm \tau_i |I_i)$. 
Anecdotal evidence points to similarities between the encoder implementation and explicit latent variables~\cite{Edelberg_VAEs}.

The optimization objective for the encoder and decoder networks is the
standard VAE variational lower bound of the model evidence, 
which includes the reconstruction error as the mean squared error between the
reconstructed image and the input image.
We note that the encoder and decoder networks are optimized together for each dataset; neither is pretrained on other datasets.

The original CryoDRGN implementation was demonstrated with known precomputed viewing 
directions. Followup work~\cite{zhong_cryodrgn2_2021} demonstrated a hybrid 
use of explicit latent variables for viewing directions and translations, and 
an encoder for the conformation variables. 
Related work on CryoFIRE~\cite{levy_amortized_2022} 
uses an encoder that outputs both conformation and pose variables using
amortized inference.

\subsection{Adding Model Constraints}

The functions used to express hyper-molecules in \cite{lederman_continuously_2017,lederman_hyper-molecules_2020} and the deep learning version in \cite{zhong_reconstructing_2019} are very generic in their ability to express a wide variety of changing volumes; including ones that are not physically plausible. 
In \cite{lederman_hyper-molecules_2020}, it is pointed out that complex heterogeneity presents an information problem. This can be illustrated by looking at a simplified discretized case.
Suppose that a good reconstruction of a volume requires $10^5$ particle images. Now suppose that the molecule has a flexible region and that the flexibility can be adequately represented using $10$ states, each requiring $10^5$ particle images for a good reconstruction (we will see in the discussion that continuous models may require fewer particle images, but the principle is similar). 
For convenience, let us assume that the particle images are conveniently equally distributed among the different states and that we are able to associate each image to the correct state.
Now suppose that there are $2$ such flexible regions; this means that there are $100$ combinations of states of the two regions, each requiring $10^5$ particle images. The number of particle images grows exponentially fast with the complexity and number of regions (if not mitigated somehow). 
\cite{lederman_hyper-molecules_2020} proposes to enforce structure in the models that capture any of the physical properties that can be assumed about the molecule; the example presented there is explicit decomposition of the volume into regions that are allowed move independently, and it is argued that the number of coefficients describing the problem (loosely reflecting on the number of particle images) grows linearly (or even remains constant) as the number of regions grows.
We note that in contrast to the multi-body approach to decomposition of the volume (Section \ref{sec:multibody}), the components in the example in \cite{lederman_hyper-molecules_2020} are not rigid.

The following sections discuss several different approaches that enforce models on hyper-molecules.

\subsection{GMM Model}
\label{sec:gmm}

A Gaussian mixture model (GMM) was introduced in~\cite{chen_deep_2021} as a 
different model for expressing $V(\bm \tau, \mathbf{r})$\footnote{
    The code is distributed as part of the EMAN2 software at
    \url{https://blake.bcm.edu/emanwiki/EMAN2/e2gmm}. 
}. In GMM models, the 
volume is represented as a sum of $N$ Gaussians centered at spatial coordinates
$\left\{\mathbf{r}_j(\bm \tau)\right\}_{j=1}^N \subset \mathbb{R}^3$,
which vary continuously with the conformation variable~$\bm \tau$:
\begin{equation}
    V(\bm \tau, \mathbf{r}) = 
    \sum_{j=1}^N a_j(\bm \tau) e^{
        -\frac{
            |\mathbf{r} - \mathbf{r}_j(\bm \tau)|^2
        }{
            2\sigma_j(\bm \tau)^2
        }
    },
    \label{eq: gmm nn}
\end{equation}
where $\{a_j(\bm \tau)\}_{j=1}^N, \{\sigma_j(\bm \tau)\}_{j=1}^N \subset \mathbb{R}_{+}$ 
are the amplitudes and widths of the Gaussians respectively 
and are also dependent on the conformation variable $\bm \tau$.

The GMM model in \cite{chen_deep_2021} is based an autoencoder architecture like CryoDRGN. 
The decoder in this model takes the latent conformation variable $\bm \tau$ (with a default dimension four) and produces a list of centers $\left\{\mathbf{r}_j(\bm \tau)\right\}_{j=1}^N$ that can then be used to evaluate the tomographic projection for any viewing direction.
The encoder computes the latent conformation variable $\bm \tau$ for any given images (with notable differences that we omit here from the CryoDRGN encoder).

The training of the neural network is performed in two steps: 
the encoder is first trained to match a pre-computed neutral
structure corresponding to $\bm \tau = \bm 0$, and then both
the encoder and the decoder are trained to match the variability 
in the data. 
The second step of the training is similar to that of CryoDRGN.
variable $\phi_i$.

\begin{remark}
    The NMA representation using pseudoatoms with locations given 
    in~\eqref{eq:linearatom} is a special case of the GMM representation 
    in~\eqref{eq: gmm nn} (but not the specific algorithm discussed in this section). 
    We can see this by letting 
    $a_j(\bm \tau) = a_0$ and $\sigma_j(\bm \tau) = \sigma_0$ 
    for all $j = 1, \ldots, N$, where $a_0$ and $\sigma_0$ are the fixed values
    of the amplitudes and widths of the pseudoatoms in~\eqref{eq:linearatom}
    and by letting
    \begin{equation}
        \mathbf{r}_j(\bm \tau_m) =  
        \mathbf{r}_j^0 + \sum_{k=1}^K b_k(\bm \tau_m) \Delta_{k_j},
    \end{equation}
    where $\mathbf{r}_j^0 \in \mathbb{R}^3$ is the vector of coordinates 
    corresponding to the $j$-th pseudoatom in the reference structure, 
    $\Delta_{k_j} \in \mathbb{R}^3$ contains the entries of the normal mode 
    $\Delta_k$ corresponding to the $j$-th pseudoatom, 
    and the normal mode coefficients $b_k^m := b_k(\bm \tau_m)$ are 
    functions of the conformation variable $\bm \tau_m$.
\end{remark}

\subsection{GMM with Folding Constraints}
\label{sec:gmm folding}

The methods described thus far make limited use of what we know about the molecules imaged in the experiment. 
Biomolecules are a folded chain of amino acids, and in many cases one has information about the chain and even some approximation of the folded structure.
This information can be incorporated into a more nuanced model.

AtomVAE~\cite{rosenbaum_inferring_2021}~\footnote{
    The survey in \cite{donnat_deep_2022} refers to this method as AtomVAE.
} models the molecule at the atomics level, expressing $f(\bm \tau, \mathbf{r})$ 
as a GMM model where each atom (other than hydrogen) is represented by a Gaussian. The architecture of AtomVAE is also based on a VAE approach,  
where the encoder consists of three networks: one that takes
a particle image and encodes it into a general latent vector, 
one that takes the general latent vector and encodes the conformational 
landscape, and one that take the general latent vector and encodes
the pose. The decoder takes the sampled latent conformation vector
and the pose as inputs and outputs the translations with respect to a neutral 
structure, previously obtained using homogeneous reconstruction 
(or purely computational methods).

CryoFold~\cite{zhong_exploring_2021} models the molecule at the residue level, 
expressing $f(\bm \tau, \mathbf{r})$ as a GMM model where each amino acid is 
represented by two Gaussians, one for the backbone and one for the side chain. 
It assumes known, fixed, viewing directions and uses a VAE architecture 
similar to that of CryoDRGN and outputs the translations of the Gaussians with 
respect to a neutral structure, as done by AtomVAE.

In both methods, structural constraints related to the positions of the
atoms in the backbone and the in side chains are specified in the loss function, 
in addition to the error in the reconstructed images.
They have been demonstrated on synthetic data and they require 
a good initialization with a reasonable reference structure.

\subsection{Deformation Based Models}
\label{sec:deformation}

It is appropriate to model many cases of continuous heterogeneity as 
deformations of a reference volume; this may exclude the binding and unbinding 
of subunits or ligands, such as elongation factors on ribosomes. 
Given a reference volume $V_0(\mathbf{r})$, a hyper-molecule description of 
the deformed volume can loosely be formulated as 
$$
    V(\bm \tau, \mathbf{r})=V_0(g[\bm \tau]^{-1}(\mathbf{r})),
$$ 
where $g[\bm \tau]$ expresses the deformation of the reference to the 
conformation $\bm \tau$ (a more nuanced technical description may also 
account for the Jacobian of the deformation). 
Deformation based models attempt to fit both the reference 
volume $V_0(\mathbf{r})$ and the deformations $g(\bm \tau)$.

Deformation functions based on Zernike polynomials have been used 
in~\cite{calero2021continuous,herreros_approximating_2021}, which combine 
ideas from manifolds of volumes described in Section~\ref{sec:manifoldvol}.
3DFlex\footnote{
    3DFlex is part of the CryoSPARC software package 
    (link in footnote~\ref{fn:relioncryosparc}).
}~\cite{punjani_3d_flex_2021} uses a deep neural network that produces a 
deformation field generator $g$ as a function of the latent conformation 
variable $\bm \tau$. 3DFlex optimizes the parameters of the deformation 
generator network, the canonical volume, and the latent conformation variable 
of each particle image in a maximum-likelihood framework. The viewing 
directions and the CTF parameters are assumed to be known, but can be 
further refined by 3DFlex.

Follow up work on the GMM method in~\cite{chen_deep_2021} is presented 
in~\cite{chen_integrating_2022}, which combines physical constraints and 
deformation models in a more scalable version of the original GMM algorithm.

\section{Bypassing the Estimation of Latent Variables}
\label{sec:matching dist}

Let us assume that the particle images are generated according to a classic 
generative model, where latent variables such as conformation, viewing 
direction, translation (and possibly CTF parameters) are selected at random 
and used to generate a particle image along the lines 
of Equation~\eqref{eq: image formation model}. 
Let us further assume that the distributions of these latent variables and the 
noise are known. In theory, this would give us a distribution of particle 
images that is conditional on the hyper-molecule. 

It was shown in \cite{panaretos2009random,gupta_cryogan_2021,bendory2017bispectrum,bandeira2020optimal} 
that a given distribution of particle images can be theoretically traced back 
to a unique volume (up to trivial symmetries). For simplicity, we 
limit the description in this section to the problem of homogeneous 
reconstruction when there is no conformational heterogeneity in the data 
and we will point to the way it generalizes to the heterogeneous case without 
delving into details.

One of the methods to compute parameters from a distribution in many 
statistical problems is the method of moments, where moments of a 
distribution are computed and used to estimate the parameters of the 
distribution. Indeed, back in the 1970s, Zvi Kam found that one can compute 
moments of the distribution of observed images and use these to reconstruct 
the volume~\cite{kam1980reconstruction}. The method has been further developed 
in recent years, for example in~\cite{bendory2017bispectrum,bandeira2020optimal,perry2019sample}.
Remarkably, this can be applied directly to micrographs, without particle 
picking~\cite{bendory2019multi}. This approach, as well as the broader area of 
multi-reference alignment, has been shown to extend to the case of heterogeneity 
(e.g.,\cite{lederman_representation_2020,boumal_heterogeneous_2018}).

Recently, machine learning approaches have emerged for approximating 
distributions. CryoGAN~\cite{gupta_cryogan_2021} is a machine learning approach 
that directly finds a volume that generates the distribution of observed 
particle images. We note that, in practice, the data consists of a large but 
finite number of samples from the distribution, not the distribution itself. 
Considering the many latent variables (viewing direction, shifts, CTF), it 
might be surprising that this is enough for CryoGAN to recover a volume, but 
the idea has been demonstrated in~\cite{gupta_cryogan_2021}. The method has 
been extended to the case of heterogeneity in~\cite{gupta_multi-cryogan_2020}.

\section{Discussion and Perspectives}
\label{sec:discussion}

Having surveyed the main approaches to continuous heterogeneity analysis,
in this section we discuss some of the broader in these approaches 
pose.

\subsection{Reading the Output: Explicit Models vs. Reconstruction from Images}\label{sec:reconstruct}

Given the output of a continuous heterogeneous reconstruction algorithm, there 
are generally two approaches to obtain the volume corresponding to a conformation
at a specific value of the conformation variable $\bm \tau = \bm \tau^*$. 

One approach involves selecting all the particle images $\{I_i\}$ that have been
assigned values $\{ {\bm \tau_i}\}$ that are within a certain distance from 
$\bm \tau^*$ and feed them to a homogeneous reconstruction algorithm. 
This produces one volume that represents that specific region of the 
conformation space.  
Alternatively, one can directly use the explicit model generated by the 
algorithm. For example, the hyper-molecule models described in 
Section~\ref{sec:hyper} have an internal representation of each state,
which allows the user to generate the volume $V(\bm \tau^*, \cdot)$. 
Similarly, equation~\eqref{eq:linearvol:modes} gives a way to visualize
principal volumes.

Some of the methods presented in this survey rely on homogeneous 
reconstruction to generate volumes with different conformations, as they lack
an explicit representation of the volumes (the manifold of images methods 
in Section~\ref{sec:manifoldEM} are an example).
However, most of the other methods lend themselves to both approaches to 
generating heterogeneous volumes. In this case, the homogeneous reconstructions
from subsets of the particle images can be used as a form of validation, 
as the final volumes are reconstructed directly from the data: if the volumes
appear to be high-resolution and biologically plausible, users tend to accept
them as a successful run of the algorithm. Such a result is especially convincing
when generating volumes directly from the model would result in volumes that
are relatively low-resolution due to model constraints such as a limited
number of principal volumes computed in the principal volume approach or a 
small number of Gaussians in the GMM models. Furthermore, as models get more
elaborate (e.g., atomic models), there are concerns that the results present
artifacts that reflect bias and error in the model, which are somewhat mitigated 
if one obtains homogeneous reconstructions from subsets of the data.

That being said, visualization based on models can provide higher-resolution
volumes, which reflects the advantages of continuous models over discrete
reconstructions. As an illustrative example, consider a molecule with one
flexible region that requires $10$ separate discrete models to be fully 
captured in the reconstruction, for which we are given $10^6$ particle images. 
If we restrict our attention to the rigid part of the volume, applying one
of the hyper-molecule algorithms would use all $10^6$ particle images for
the reconstruction of this region, while selecting subsets of $10^5$ particle 
images for homogeneous reconstruction of one of the conformations leads to only
this subset of the data being used to reconstruct the rigid part, which
would otherwise benefit from the full dataset. Modeling considerations and the
ability to refine the viewing directions taking into account flexible regions
extends this argument to other less obvious cases.

\subsection{Continuous vs. Discrete Models}\label{sec:contVsisc}

Continuous functions can be approximated by a sufficiently large number of 
discrete samples, and therefore it is compelling to deduce that traditional
discrete 3D classification is a good discretization of continuous 
heterogeneity. In this section, we point to several advantageous features
of continuous models. 

First, in contrast to discrete heterogeneity analysis, continuous heterogeneity 
analysis does not require the user to specify the number of classes to be 
reconstructed. This choice, which is not trivial when the underlying 
heterogeneity is concentrated in discrete states and influences the 
resolution of the inferred model, is particularly difficult when the 
heterogeneity is inherently continuous. 
Instead, approaches for continuous heterogeneity analysis explicitly model the
continuum of states, where states that are sufficiently distinct in the 
particle images tend to naturally appear as dense ``islands'' on the conformation 
manifold.

Second, methods for continuous heterogeneity are able to use all the available
data to produce higher resolution volumes in any conformation, while discrete
heterogeneity approaches only use a subset of the data for a particular 
conformation. The example given in Section~\ref{sec:reconstruct} illustrates 
this aspect: while the rigid part of a volume can benefit from the full dataset, 
3D classification methods only use the particle images corresponding to one
specific conformation to reconstruct the volume, including the rigid part, 
leading to lower resolution than possible.
Here, one could correctly argue that the rigid region can be reconstructed using
homogeneous reconstruction separately from the heterogeneous region, as it is 
done in multi-body refinement (Section~\ref{sec:multibody}). However, 
continuous heterogeneity analysis may still perform better, as it can update 
the viewing directions estimations (potentially yielding better estimation of the
conformation variable) and it models the continuum of states more accurately 
than discrete models, which do not have a notion of ``neighboring states''.
For example, the deformation model described in Section~\ref{sec:deformation}
severely restricts the space of possible conformations to a (limited) 
deformation of a reference volume, and therefore all the particle images 
from all conformations contribute high-resolution information to the 
reference volume.

Third, an implication of the previous argument is that continuous heterogeneity 
analysis has an advantage in analyzing rare conformations using the continuity 
between conformations. Using the deformation model as an example, when the 
deformation assumption is applicable, this model constructs a high resolution
reference volume based on the highly populated conformations and uses the
particle images from a rare conformation effectively to determine the path of 
deformation of the reference volume. This leads to a high resolution 
approximation of the rare conformation as the correct deformation of the 
reference volume.

One last argument regards the advantage of continuous heterogeneity analysis
over multi-body analysis (Section~\ref{sec:multibody}). While it has been 
implemented as an extension of traditional reconstruction algorithms,
multi-body analysis can also be seen as a special case of the hyper-molecule 
model. Specifically, is has proven to be very 
effective in the analysis of rigid components moving with respect to each 
other, but it falls short when such structural assumptions do not hold and
it fails to resolve the interface regions between the rigid components, issues
which are addressed in the hyper-molecule models for examples.

Finally, it is worth noting that high resolution homogeneous reconstruction and 
traditional 3D classification as well as multi-body analysis are available in 
mature software and are well understood by practitioners. On the other hand,
practitioners still face the dilemma of which continuous heterogeneity software, 
model assumptions and parameters to use in continuous heterogeneity analysis, 
so it is not the aim of this article to argue that the problem has been solved 
or that higher resolution is always easily attainable, but to highlight conceptual 
advantages on which these algorithms increasingly capitalize.

\subsection{Outliers, Junk, and Rare Conformations}

Practitioners have long used 2D and 3D classification not only to identify 
actual conformations, but also to identify ``junk'': particles that produce
low quality volumes and objects that were incorrectly picked as particles are
discarded. One of the caveats in this procedure is that it is difficult to 
determine whether high quality particle images are also discarded. 
Perhaps most importantly, it is plausible that particle images from rare 
conformations are also discarded.

Preliminary informal evidence suggests that an analogous procedure is 
applicable to different algorithms for continuous heterogeneity analysis: 
junk particles appear to be ``pushed'' out to lower density regions of the 
conformation space, reducing their influence on the ``valid'' regions. 
It is therefore conceivable that some of the continuous heterogeneity analysis 
algorithms strike a better balance between using rare-conformation particle 
images and discarding junk.

\subsection{Interpretation of Output}

The output of continuous heterogeneity analysis varies considerably between 
different methods. However, there are two main families of outputs. 
One is a model of the space of molecular structures (e.g., the principal 
volume expansion of  $V_m$ in~\eqref{eq:linearvol}, the hyper-molecule function 
$V(\bm \tau, \mathbf{r})$ in~\eqref{eq:hypermolecule:orthogonal}), which does 
not exist in the image manifold approach in Section~\ref{sec:manifoldEM}.
The other family of outputs consists of the estimates of the conformation
variable $\bm \tau$ for each particle image, which does not exist in the 
methods in Section~\ref{sec:matching dist}.  

From both types of outputs, one can obtain a low dimensional representation
of the latent conformation space of $\bm \tau$, where it is common to try to 
identify physically meaningful clusters or manifolds. 
For example, in~\cite{dashti_trajectories_2014}, the density of the particles 
in the latent space is related to the energy landscape of the ribosome.
In other work, various dimensionality reduction and clustering algorithms are 
applied to the latent space to identify dominating conformations. 
We note, however, that the latent conformation variable does not immediately 
correspond to a physically meaningful and interpretable quantity, it is not 
necessarily unique even up to trivial symmetries and can in principle be 
arbitrarily deformed (for some motivating examples, see~\cite{toader_remarks_2022}). 
Different algorithms and even the same algorithm with a different random seed 
can produce different latent space representations. Similarly, different 
post-processing dimensionality reduction and clustering algorithms (and again, 
even different random seeds) can produce different results.
In cases of relatively simple heterogeneity, it is plausible that an expert 
user would be able to probe structures in the latent space together with the 
models or reconstructions associated with them in order to understand the 
mechanisms that it captures. However, complex heterogeneity and more automated 
tools may require more work on the interpretation of the latent space. 

Mathematically, this is a problem related to the manifold metric as transformed 
by the measurement and reconstruction process, and while it has been studied 
in other fields
\cite{talmon_empirical_2013, meila_metric_nodate, schwartz_intrinsic_2019, bertalan_transformations_2021},
it still requires further investigation in a cryo-EM context. Moreover, 
establishing an appropriate metric for the conformation space that is 
physically meaningful is a problem that has not been explored yet.

\subsection{Validation}

Validation has always been a nuanced problem in cryo-EM. 
A common sanity check for homogeneous reconstruction is the Fourier shell 
correlation (FSC) procedure~\cite{henderson_outcome_2012, scheres_prevention_2012}: the data is split into two subsets and two distinct
reconstructions are obtained from them. The correlations between the two volumes
are computed for each frequency radius and the resolution of the final 
reconstruction is given by the frequency radius where the FSC curve drops below 
a specific value (usually 0.143). Above this resolution, the reconstruction is 
considered to overfit the noise.
In addition, a qualitative procedure is for an expert to evaluate and confirm 
the plausibility of the reconstructed molecule.

There is no clear solution for validating results of continuous heterogeneity 
analysis. To illustrate the difficulty, consider two hyper-molecules that are 
generated by one method from two separate subsets of the data. 
Even if the solutions are equivalent, they can assign different values 
of $\bm \tau$ to the same conformation, in which case it is not clear 
how to compare the two outputs.
One idea proposed in~\cite{punjani_3d_flex_2021} for deformation models 
(Section \ref{sec:deformation}) is to compute the FSC between two reference 
models that the algorithm computes for two different subsets of the data. 
However, this idea only applies to deformation models, since other methods do 
not necessarily compute a reference volume.

Given the wide variety of families of ideas for analysis of continuous 
heterogeneity and the different types of output they produce, it is even more 
challenging to find metrics that are applicable to all methods. 
Beyond the evaluation of outputs for individual problems, there is a desire to 
determine which tools perform best, which reinforces the need for evaluation 
metrics and benchmark datasets, something that is not currently available. 
Having said this, given the many forms of heterogeneity in applications and the 
fundamental differences between the algorithms for heterogeneity analysis, 
one should be cautious that the metrics of choice and the benchmark datasets 
used do not exclude some of the existing and new ideas.

\section{Conclusions}

In this survey, we covered a broad range of methods for analyzing continuous 
heterogeneity in cryo-EM. In addition, we highlighted the advantages 
of continuous heterogeneity analysis and the way they circumvent some of the
tradeoffs that practitioners encounter in traditional methods involving 
homogeneous reconstruction, discrete heterogeneous reconstruction and 
multi-body analysis.

As evident in this survey, many different approaches have been proposed in 
recent years. Most of these ideas are still in active development and it is 
likely that new ideas will emerge in the coming years. Some of the algorithms, 
like the various members of the non-linear models in Section~\ref{sec:hyper}, 
share conceptual similarities that make it plausible that we will see 
convergent software packages that offer multiple and customizable models,
as envisioned in~\cite{lederman_hyper-molecules_2020}. 
At the same time, there are significant differences between the methods, 
making different approaches well-suited for different problems.

One of the open problems in this area is the validation of the results of the 
algorithms. A related difficulty is in comparing different algorithms in the 
absence of good metrics and benchmarks; indeed, the different approaches differ 
even in the form of output they produce. A silver lining in the large number of 
fundamentally different algorithms that are becoming available is that one could 
conceivably apply several different algorithms to the same dataset and examine 
how well their conclusions agree qualitatively, which could serve as a temporary 
form of validation while the area matures.

Finally, we have we witnessed in recent years the immense success of protein 
structure prediction algorithms such as AlphaFold~\cite{jumper_highly_2021} 
and RoseTTAfold~\cite{baek2021RoseTTAfold}. While these algorithms are not 
currently applicable to continuous heterogeneity (and, arguably, neither to 
discrete heterogeneity), it is plausible that they will evolve in this 
direction as a next step in complexity. 
Conversely, as cryo-EM datasets become more massive and broader in scope,  
incorporating physical models as priors or constraints obtained from structure 
prediction software will become a necessity. Therefore, there is an opportunity 
for continuous heterogeneity analysis methods to bridge the gap between such 
software and the massive experimental datasets. Preliminary examples of this 
general direction can be found in~\cite{rosenbaum_inferring_2021}, which 
integrates insights from AlphaFold, and in~\cite{giraldo-barreto_bayesian_2021}.

While we do not discuss cryo-electron tomography (cryo-ET) explicitly in this survey, 
many of the approaches to continuous heterogeneity in cryo-EM can be generalized to cryo-ET.
One of the complications in cryo-ET is that the targets of the analysis are typically measured in situ, and do not float freely as in typical cryo-EM.
The continuous heterogeneity model may capture heterogeneity in the environment instead of heterogeneity in the target (which may or may not be a desirable result, depending on the application).

\section*{Acknowledgments}

The authors would like to thank David Silva Sanchez for his help.

The work is supported in part by NIH grants R01GM136780, R01NS02501 and AFSOR FA9550-21-1-0317.

\printbibliography

\end{document}